\documentclass{article}
\newcommand{\ie}{\mbox{i.\,e.\,\ }}
\newcommand{\iec}{\mbox{i.\,e.\,}}

\newcommand{\egc}{\mbox{e.\,g.\,}}

\newcommand{\dr}[1]{\ensuremath{\mathrm{d} #1\,}}
\newcommand{\mc}[1]{\ensuremath{\mathcal{#1}}}

\newcommand{\ddt}{\ensuremath{\frac{\dr{}}{\dr{t}}}}

\newcommand{\op}[1]{\ensuremath{\widehat{\textsf{\ensuremath{#1}}}}}
\newcommand{\opad}[1]{\ensuremath{\op{#1}^{\dagger}}}
\newcommand{\id}{\op{\mathsf{1}}}

\newcommand{\comm}[2]{\ensuremath{\left[ #1 , #2 \right]}} 
\newcommand{\tr}{\textsf{Tr}}

\newcommand{\be}{\begin{equation}}
\newcommand{\ee}{\end{equation}}
\newcommand{\e}[1]{\mathrm{e}^{#1}}

\usepackage{chicago}
\usepackage{amssymb}

\newcommand{\superop}[1]{\mathbf{#1}}

\begin{document}

\title{Quantum systems other than the Universe}
\author{David Wallace\thanks{Department of History and Philosophy of Science / Department of Philosophy, University of Pittsburgh; \texttt{david.wallace@pitt.edu}}}
\maketitle

\begin{abstract}
How should we interpret physical theories, and especially quantum theory, if we drop the assumption that we should treat it as an exact description of the whole Universe? I expound and develop the claim that physics is about the study of autonomous, but not necessarily isolated, dynamical systems, and that when applied to quantum mechanics this entails that in general we should take quantum systems as having mixed states and non-unitary dynamics. I argue that nonetheless unitary dynamics continues to have a special place in physics, via the empirically-well-supported reductionist principles that non-unitarity is to be explained by restriction to a subsystem of a larger unitary system and that microscopic physics is governed by unitary and largely known dynamics.  I contrast this position with the `Open Systems View' advocated recently by Michael Cuffaro and Stephan Hartmann.
\end{abstract}

\section{Introduction: Physics as the study of autonomous systems}\label{introduction}

Quantum theory is often described as our most fundamental theory of physics, and as such, the task of understanding what quantum theory tells us about the world is often taken to be the task of understanding what the world would be like if quantum theory in fact described it completely. To understand quantum theory in this way commits us (here I follow~\cite{wallace-isolated-1}) to two assumptions: the \emph{Fundamentality Assumption}, that the theory is \emph{exactly true}, and the \emph{Cosmological Assumption}, that the theory describes \emph{the whole of reality}.\footnote{For evidence that the two assumptions are widespread, see \cite[section 1]{wallace-isolated-1}. There are exceptions: notably, the conjunction of the two assumptions gives us something close to Ruetsche's~\citeyear{ruetschebook} notion of a `pristine interpretation' of a quantum theory, and the alternative that she proposes is close to what I propose here. See also~\citeN{williamseffectiverealism}, who proposes something similar in the context of effective field theories, though focussed more on rejection of fundamentality than of cosmology.} Of course, no-one literally thinks that any particular quantum theory --- or indeed any theory we currently possess --- actually satisfies either assumption, but still: much, perhaps most, interpretative work on quantum theory and indeed on theories of physics in general takes them as a starting point. 

At the risk of clich\'{e} (cf \citeNP{quinedogma}), we might call these two \emph{dogmas} of interpretation --- dogmas that, once made explicit, are very hard to argue for. A moment's observation of physics shows that hardly ever do physicists seek to model the whole universe, and hardly ever do they seek to model every last degree of freedom of even a finite system. What actually happens in physics, in almost all cases, is that a physical theory is used to model some subsystem of the universe, and to model that subsystem only on certain scales. (Even in general relativity, where our models are called `spacetimes' and it is only natural to think of them as modelling entire Universes, in practice the vast majority of the time we are actually using them to model isolated subsystems of a larger Universe. Even when applied to cosmology, our models usually describe patches of a larger cosmos, and normally track only large-scale features of the geometry and matter distribution.)

Of course, it remains logically possible that our understanding of physics modelling practice is somehow deriviative on the interpretation of theories under the Fundamentality and Cosmological assumptions. But there seems little positive reason so to interpret them: what we actually have evidence for is the efficacy of physics in modelling scale-dependent features of subsystems, and it seems far more natural to look for a direct understanding of physical theories --- and, in particular, quantum theories --- as they are so used.

But \emph{which} subsystems are the proper concern of physics? A very natural answer --- one I (too quickly) endorsed in \citeN{wallace-isolated-1} --- is that physics models \emph{isolated} systems, systems that do not interact with any other system. But this has multiple problems. For one thing, it is not as straightforward as it might seem to say carefully what `isolated' actually means here --- the temptation is to say something like `evolving the same way as it would if it were alone in the Universe', but that is scarcely a counterfactual that we can assess empirically. For another, insofar as we can understand `isolation' it seems to apply to systems that are dynamically isolated from one another in some (usually spatial) way, and does not naturally apply when we are modelling some degrees of freedom of a system and neglecting other, perhaps smaller-scale, degrees of freedom of that same system.

We could go formal. Quantum mechanics --- at least as it is normally formulated --- is a theory with unitary dynamics, where a system evolves under the Schr\"{o}dinger equation with respect to some fixed Hamiltonian. We could \emph{define} an isolated quantum system as a system that could be so modelled. 

One problem with this definition is that unitarity does not coincide with isolation in the physical sense. Sometimes an environment interacts with a system not by introducing non-unitary dynamics but by adjusting --- \emph{renormalizing} --- the Hamiltonian of the system. But there is a much more serious problem, which also afflicts any attempt to make precise a more physical notion of isolation. The problem is that `isolated system' under any reasonable disambiguation is inadequate to capture all the systems that physics (including quantum physics) in fact succeeds in modelling. The excited atom that decays through the emission of radiation; the pollen grain undergoing Brownian motion; the friction-bound system that eventually slows to a stop\ldots all these are paradigmatically important, yet paradigmatically \emph{non}-isolated, systems. Many such cases, indeed, are paradigmatically important \emph{quantum} systems, but they are not isolated in either the substantive sense that they can be physically understood without consideration of their environment, or in the formal sense that their dynamics can be reduced to the Schr\"{o}dinger evolution of pure states.

Let me suggest a better answer to the question of physics' proper concern. What physics models is \emph{autonomous} systems, and a system is autonomous if it can be effectively modeled \emph{as} a system with a state space and a dynamics, such that many initial states of the system are possible and that the dynamics tells us what the future states are as a (possibly stochastic) function of the present and past states. In particular, a system is an autonomous \emph{quantum} system if it can be effectively modelled as a system with a Hilbert space and a dynamics for (not necessarily pure) quantum states on that Hilbert space.

All the systems I mention above are autonomous in this sense --- of course various things must hold of the environment for the system to be autonomous, and broad features of the environment (the temperature of the fluid in which the pollen grain moves, for instance) determine the dynamical parameters, but nonetheless it is possible to state, and test, a dynamical equation for the system's degrees of freedom, without regard for the \emph{details} of the state of the environment. 

Autonomy is neither trivial nor automatic: most subsystems of a complex interacting system will \emph{not} be autonomous. (For instance, consider three bodies of similar masses interacting gravitationally: the three bodies form a complex interacting system and no two bodies can be modelled effectively, even stochastically, without tracking the third.) Nor need science be concerned only with autonomous systems. A non-autonomous system can be thought of as requiring not only initial-time data but a constant stream of data describing the influence of the environment, data which cannot itself be modelled (at least not as part of studying this system) but is simply an input. The human body, or an organ of the body, can profitably studied this way: we want to know not what it does in isolation but what it does as a function of the way the environment impinges upon it over time.

But for the most part, \emph{physics} is concerned with autonomous systems. The primary subject matter of physics is dynamics, and non-autonomous systems have no dynamics: their behavior cannot be analyzed dynamically except as a shadow on the wall, cast by the dynamics of a larger system. 

My main goal in this paper is to develop what an account of quantum systems looks like given that it should be a theory of \emph{autonomous} systems, but not necessarily isolated systems --- still less systems that aim to model the Universe. The starting point in this account (developed in section~\ref{open-system-derivation}) is the observation that \emph{formal} features of quantum theory, when combined with appropriate choices of system and dynamics ---  imply a robust notion of autonomy that is equivalent neither to unitary dynamics nor to the physical notion of isolation. We can obtain that notion by (i) writing a formally-exact equation for the evolution of the reduced state of any subsystem of a unitarily-evolving quantum system, and (ii) identifying physical assumptions under which that formally-exact equation reduces (in some appropriate idealization) to an autonomous dynamical equation for the subsystem alone.  This analysis shows that in some sense unitary dynamics is unstable: the existence of large systems with unitary dynamics often entails the existence of smaller autonomous subsystems with non-unitary dynamics.

This observation, especially once we reject the Fundamentality and Cosmological Assumptions, might suggest that unitarity is a ladder that can be kicked away, and that we should replace it entirely with a generalized, non-unitary (perhaps also non-Markovian) notion of quantum dynamics.  I do not think this is true. Quantum physics may not be about exactly-true models of the whole Universe, but it does provide very powerful (essentially inductive) evidence for a thesis of \emph{quantum reductionism}: that non-unitarity can and should always be understood as the restriction to a subsystem of larger-scale unitary dynamics, and that larger-scale, phenomenological quantum physics is ultimately explicable microscopically in terms of specific, largely known, unitary quantum theories: non-relativistic particle mechanics and various quantum field theories. I develop this position in section~\ref{reductionism}, where I suggest four assumptions of quantum reductionism which in a way constitute a naturalistic replacement for the Fundamentality and Cosmological assumptions.

I am not the first to develop a view of quantum systems along these general lines, and indeed this paper is substantially inspired by recent pioneering work by Cuffaro and Hartmann~\citeyear{cuffarohartmann}. Their \emph{Open Systems View} rejects the primacy of unitary dynamics and aims at analyzing open quantum systems, not just isolated ones, and my second goal in this paper is to engage with their view and see how it relates to mine. In section~\ref{opensystems} I argue that their view tacitly requires the notion of autonomy which I have described above, and that once this is recognized, the main difference between our views concerns the status of quantum reductionism: at least on one interpretation, the Open Systems View is committed to the rejection of most or all of my four assumptions of quantum reductionism. As I go on to argue, in large part I think this feature of their view needs to be rejected, for a mixture of reasons: partly that it makes the very notion of `interaction' mysterious in a way which inhibits physics, partly because of the evidential support for reductionism in physics practice. But in section~\ref{arrowoftime} I identify an important --- if partial --- exception to my arguments, concerned with how \emph{irreversibility} enters into the dynamics of autonomous systems, and relate the differences between my view and the Open Systems View to unresolved issues in the philosophy of statistical mechanics.

The physics I discuss is fairly standard and for the most part I do not provide original references. An up-to-date standard monograph on open quantum systems is \cite{breuerpetruccionebook}.

\section{Open quantum systems: deriving autonomy from unitarity}\label{open-system-derivation}

Suppose that some unitarily evolving quantum system with Hilbert space $\mc{H}$ and Hamiltonian $\op{H}$ has a tensor-product decomposition into subsystems $S$ and $E$, $\mc{H}=\mc{H}_S\otimes \mc{H}_E$, and a concomitant decomposition of its Hamiltonian,
\be
\op{H}=\op{H}_S\otimes \id  + \id\otimes \op{H}_E + \sum_n \op{C}_n \otimes \op{D}_n.
\ee
If the third term in this expression vanishes, $\mc{H}_S$ and $\mc{H}_E$ will define closed systems with Hamiltonians $\op{H}_S$, $\op{H}_E$ respectively. Even in this case, of course, the embedding of these systems in a larger system means that their joint state might be entangled, so that the appropriate notion of \emph{state} for each subsystem is a mixed rather than pure state. (This already suffices to demonstrate that once we drop the Cosmological Assumption, the principle that a quantum system must be in a pure state becomes untenable; cf~\cite{wallaceirreversibility} for further development of this observation in the context of quantum statistical mechanics.)

However, the vanishing of the interaction term is not the most general circumstance in which a subsystem of a larger quantum system is autonomous. To see this,\footnote{The analysis I provide here is one of two main routes in quantum statistical mechanics to derive autonmous dynamics via scale separation. The other --- the quantum version of the so-called BBGKY hierarchy --- seeks to develop not an autonomous dynamics of subsystems but an autonomous dynamics of $N$-point correlation functions for some fixed (usually small) $N$. See \citeN[sections 2-4,10]{wallacequantumstatmech} for a conceptual discussion of this latter approach; see also  \cite[ch.1]{calzettahubook} for a comparison.} let $S_E$ be a fixed (in general mixed) state of $\mc{H}_E$, which we will require to be invariant under the self-Hamiltonian $\op{H}_E$ of $\mc{H}_E$, $[S_E,\op{H}_E]=0$. Then define the following superoperator $\superop{J}$ on the joint density operators of $\mc{H}_S\otimes \mc{H}_E$:
\be
\superop{J}\rho = \tr_E \rho \otimes S_E
\ee
where $\tr_E$ is the partial trace over $\mc{H}_E$, \ie $\tr_E\rho$ is the reduced density operator of $\mc{H}_S$. Note that $\superop{J}$ is a projection, $\superop{J}^2=\superop{J}$, projecting onto the subspace of states of form $\rho=\rho_S\otimes S_E$. (Also note that here and after, I (i) use $\mathbf{boldface}$ to label superoperators that act on the space of Hilbert-space operators; (ii) drop the `hats' on mixed quantum states, to avoid cluttering the notation.)

Following \citeN{nakajima} and \citeN{zwanzig} we can now extract an integro-differential equation for $\rho_S$, the \emph{Nakajima-Zwanzig equation}:
\begin{eqnarray}\label{exactexpression}
\ddt{\rho_S(t)}&=& - i \comm{\op{H}_R}{\rho_S(t)}+ \tr_E\int_0^t\dr{\tau}\superop{L}_H \superop{V}(\tau)(\superop{1}-\superop{J})\superop{L}_H(\rho_S(t-\tau)\otimes S_E)\nonumber \\
&+& \tr_E \superop{L}_H \superop{V}(t)(\rho(0)-  \rho_S(0)\otimes S_E)
\end{eqnarray}
where:
\begin{itemize}
\item $\op{H}_R$ is the \emph{renormalized system Hamiltonian}, 
\be
\op{H}_R = \op{H}_S + \sum_n \op{C}_n \tr(\op{D}_n S_E).
\ee
\item 
$\superop{L}_H$ is the superoperator representing the adjoint action of the Hamiltonian,
\be
\superop{L}_H\rho= -i \comm{\op{H}}{\rho}.
\ee
\item $\superop{V}(t)$ is a modified propagator,
\be
\superop{V}(t)=\e{(\superop{1}-\superop{J})\superop{L}_H t}.
\ee
\end{itemize}
Equation (\ref{exactexpression}) is exact, and so by itself tells us nothing about whether subsystem $S$ has autonomous dynamics. However, in many cases of physical interest there is \emph{separation of scales}, so that the timescale $T_E$ on which the degrees of freedom of $E$ evolve --- and, in particular, effectively equilibrate --- is much faster  than the timescale $T_S$ on which the degrees of freedom of $S$ evolve and interact with $E$. When this occurs (something that needs to be argued for in any specific case), then if we take $S_E$ to be the appropriate equilibrium state (\iec $S_E \propto \exp(-\beta \op{H}_E)$ for some inverse temperature $\beta$),
\begin{enumerate}
\item Since effective equilibration of a system means that the matrix elements of those operators describing collective degrees of freedom --- such as those describing coupling with other systems --- approach their equilibrium values, for $t\gg T_E$ we ought to be able to approximate $\rho_0$ by $\rho_S \otimes S_E$ in the third term, so that it approximately vanishes.
\item Since the second term tracks the degree to which previous states of $S$ leave a memory trace in $E$, and since such traces ought to be effectively unraveled by equilibration, the kernel in that second term ought to fall off on timescales $\sim T_E$, so that for $t\gg T_E$ we can replace the upper limit $t$ of integration with some fixed large (but sub-recurrent) time $T$.
\end{enumerate}
Equation (\ref{exactexpression}) then simplifies to
\be\label{effectiveexpression}
\ddt{\rho_S(t)}= - i \comm{\op{H}_R}{\rho_S(t)}+ \tr_E\int_0^T\dr{\tau}\superop{K}(\tau)(\rho_S(t-\tau)\otimes S_E)
\ee
where $\superop{K}(\tau)=\superop{J}\superop{L}_H \superop{V}(\tau)(\superop{1}-\superop{J})\superop{L}_H$. This is an autonomous, time-translation invariant, evolution equation for $\rho_S$. It is, however, \emph{non-Markovian}: the system has memory, and its future evolution depends on its past and not just its present state.

Systems which are \emph{sufficiently} non-Markovian are often autonomous only by courtesy. The interactions with the system's environment leave traces in the system's past state, and those traces can be used to reconstruct the environment state. (It is plausible that one could write down a wildly non-Markovian ``autonomous'' dynamics for one body in the three-body problem.) Autonomy requires that the kernel $\superop{K}(\tau)$ falls off quickly compared to the timescales on which the system evolves, so that only its \emph{recent} past is dynamically relevant. We have already assumed this, however, and so equation (\ref{effectiveexpression}) indeed describes an autonomous dynamics. 

In many cases we can further assume that the memory timescale is short enough that $\rho_S(t)$ is effectively static on that timescale, in which case the second term can be approximated as
\be
\tr_E\int_0^T\dr{\tau}\superop{K}(\tau)\rho_S(t-\tau)\otimes S_E\simeq \tr_E\int_0^T\dr{\tau}\superop{K}(\tau)(\rho_S(t)\otimes S_E).
\ee
If we then define $\superop{\Lambda}$ by
\be
\superop{\Lambda} \rho_S = \tr_E \int_0^T \dr{\tau}\superop{K}(\tau) (\rho_S \otimes S_E)
\ee
we have extracted a \emph{Markovian}, but in general non-unitary, dynamical equation for $\rho_S$:
\be
\ddt{\rho_S(t)}=- i \comm{\op{H}_R}{\rho_S(t)} + \superop{\Lambda} \rho_S(t).
\ee
Fairly general results from the theory of open quantum systems (see, \egc, \citeNP{nielsenchuang}) tell us that in this case, $\superop{\Lambda}$ must be expressible in Lindblad form: that is, there must exist operators $\op{L}_i$ such that
\be
\superop{\Lambda} \rho = \sum_i \left( \op{L}_i \rho \opad{L}_i+ \frac{1}{2}\left( \op{L}_i\opad{L}_i \rho + \rho \op{L}_i \opad{L}_i\right)\right).
\ee

To sum up, we have a hierarchy of possibilities for a quantum system. At the bottom of the hierarchy, the system might be a `system' only by courtesy, inextricably dynamically interwoven with its environment and studyable only in conjunction with it. Alternately, it might be autonomous, governed by a dynamics which does not depend on the fine details of the environment (in the case we have considered, it depends only on the temperature) and which, if it has memory effects, has those effects only on timescales fairly short compared to its dynamical timescale. Above that in the hierarchy, it might be Markovian, with the memory effects idealizable into an instantaneous dynamics. And if those dissipative effects are entirely absent or can be idealized as such, the system is unitary.

Note that (as anticipated in section~\ref{introduction}) `isolated' does not appear in this hierarchy, and indeed is not straightforwardly definable in formal terms. It cannot be identified with `unitary': as we have seen, even if dissipative effects of other systems can be neglected, other systems' influence can appear as a renormalization. And this is common in physics: the famous renormalizations of quantum field theory have exactly this form, for instance, where the effects of high-energy physics manifest at lower energies not through any violation of unitarity, but through the renormalization of the parameters in the unitary dynamics. (See \cite{boyanovsky} for an explicit presentation of quantum-field-theoretic renormalization in this form.)

\section{Four dogmas of quantum reductionism}\label{reductionism}

We have seen that it is not stable to suppose that quantum systems invariably or even usually have unitary dynamics. Once we realize that quantum physics, like physics generally, is concerned with autonomous subsystems of the universe and not simply with the universe as a whole or those of its subsystems that can be treated as entirely isolated, we have to recognize that quantum systems generally must be allowed to have mixed states and that quantum dynamics need not be Markovian and, even when Markovian, need not be unitary.

Once we recognize this, we could in principle adopt `quantum pluralism': the belief that there is nothing particularly special about unitary dynamics (or pure states) and that the right way to do quantum mechanics is simply to discover, system by system, what that system's dynamics actually are, free of any prejudice that they must be unitary. Quantum pluralism is a fundamentally disunified quantum metaphysics, very reminiscent of Nancy Cartwright's ``dappled world'' metaphysics, in which we abandon any pretense for universality of any dynamical laws and work system-by-system, and quite closely related to Cuffaro and Hartmann's `Open Systems View' (to which I will return in section~\ref{opensystems}).

But the vast majority of practicing physicists do not practice quantum pluralism, and I think they are correct not to do so. Physics practice is overwhelmingly committed to what I call \emph{quantum reductionism}, a position that is encapsulated by four principles --- four dogmas of quantum reductionism, if you like. The principles can be understood both methodologically (this is how, as practicing physicists, we \emph{should} explore the physics of quantum systems) and metaphysically (this is, in fact, the \emph{truth} about quantum systems). I should be clear that I use `dogma' playfully here: in my view the four `dogmas' of quantum reductionism are fully justified methodologically and probably largely correct metaphysically, and indeed I offer them as a replacement for the two dogmas we have already encountered (the Fundamentality and Cosmological assumptions).

The first principle is that \emph{non-unitarity implies interaction}. A unitarily evolving quantum system may or may not require us to consider its interactions with other systems in order to explain its dynamics (recall renormalization), but if a system is non-unitarily evolving it is always doing so because of its interactions with other systems --- or at least, denying this is an absolute last resort. More specifically: the presence of memory terms always indicates that some additional and unmodelled degrees of freedom are interacting with the system; even in the absence of memory terms, non-unitarity means that the system is becoming entangled with some other system.

The second principle is that \emph{interaction is derived from dilation}. By a \emph{dilation} of an autonomous quantum system, I mean the inclusion of that system in a larger autonomous quantum system, such that the dynamics of the original system can be derived from the dynamics of the larger system by the means sketched in section~\ref{open-system-derivation}.

Section \ref{open-system-derivation} made only minor use of the unitarity of the larger system. The superoperator $\superop{L}_H$ defined the dynamics of that system, and I took $\superop{L}_H\rho=-i[\op{H},\rho]$, but the derivation of equation (\ref{exactexpression}) would go through largely unchanged if $\superop{L}_H$ were itself a non-unitary dynamics. (The main difference is that the first term in the expression would become non-unitary.) But in practice this is not how dilations are normally done in physics, and this is the third principle: that \emph{the dilations that explain interaction are to larger unitary systems}. More precisely: they are to systems that can be idealized as unitary at least for the purposes of explaining the subsystem's dynamics: perhaps their own dynamics, modelled at a sufficiently high level of precision, requires non-unitary terms, but those terms make no significant contribution to the subsystem's autonomous dynamics. (If they did, then by the third principle we did not dilate far enough --- but the third principle commits us not just to some in-principle possibility of dilating unitary but to the practical strategy of modelling interaction that way.)

At a purely formal level, the dilational assumptions in the second and third principle are almost empty: Stinespring's~\citeyear{stinespring} dilation theorem (see, e.g., \cite{nielsenchuang} for a modern presentation) says that (under weak additional assumptions; cf \cite[section 2.2.2]{cuffarohartmann}) any well-behaved linear dynamics for a quantum system is the restriction to that system of \emph{some} unitary dynamics on \emph{some} subsystem. The second and third assumptions get their bite from the requirement that the dilation needs to be not just to a fictional system but to some concretely specified system whose dynamics can themselves actually be observed and measured.

Indeed, in physics practice it is ultimately insufficient merely to posit the dynamics of that larger system, even if those posits are observationally confirmed. The fourth principle is that the dynamics of any autonomous quantum system should ultimately be derived, via restriction to subsystems and/or the averaging over microscopic degrees of freedom (insofar as these are distinct), and under appropriate restrictions on a system's state (e.g., fixing temperature or a regime of symmetry breaking), from a short list of ``microscopic'' --- and invariably unitary --- quantum theories. The main entries on the list are:
\begin{enumerate}
\item Non-relativistic quantum mechanics: The nonrelativistic physics of nuclei and electrons, interacting via Coulomb coupling, occasionally also by Newtonian gravitational coupling, and by coupling to background electric, magnetic, and gravitatational fields.
\item Relativistic quantum mechanics: The relativistic generalization of (1), under the self-consistency assumption that interparticle interactions are gentle enough not to lead to particle creation.
\item Quantum optics, the hybrid theory of the quantized electromagnetic field coupled to nonrelativistic nuclei and electrons.
\item Quantum electrodynamics, describing relativistic electrons and nuclei (the latter treated as pointlike) interacting with the quantized electromagnetic field.
\item The Standard Model of Particle Physics, along with its various effective-field-theory approximations (which in fact include (1)-(4).)
\item Low-energy quantum gravity, the effective field theory of gravitation coupled to quantum matter. (Strictly speaking this is the only item we need on the list: (5), and hence (1)-(4), are derivable from it.)
\end{enumerate}
The fourth principle is not inviolate, and is routinely violated in places where `new physics' is expected or at least not ruled out: examples include dark matter, beyond-standard-model particle physics, and the observationally underconstrained mechanisms of CP violation and neutrino mass. But the positing of new physics (like fundamental non-unitarity) is normally a last resort, and this is especially true in terrestrial, low-energy situations --- positing a new particle to explain why some high-temperature superconductor works, for instance, would require extremely powerful evidence and extensive argument as to why that particle had not shown up in other contexts.

Why think these four principles in fact constitute standard practice in physics? A really systematic demonstration would require quantitative methods well beyond the scope of this paper, but I will motivate it through two examples. The first and most important is non-equilibrium quantum statistical mechanics, where pretty much any derivation I know adheres to them. The name of the game is to derive equations --- complete with quantitatively determined parameters --- for collective degrees of freedom such as the spatially-averaged density of a fluid or a particle interacting with an environment. Invariably the derivation begins with a larger system whose dynamics are taken to be unitary; often those dynamics are taken as determined by the specific forms found in microphysical theories  non-relativistic quantum mechanics or quantum optics (or, for more exotic applications, relativistic field theory), and where they are not, we are given at least qualitative reasons to think their general form is consistent with  those microphysical theories (or else the theory is specifically described as  a toy model, intended to illustrate some general principle rather than describe some specific system). The distinction between `phenomenological' and `micro-based' equations, central in statistical mechanics, is exactly the distinction between equations posited simply to explain the data and equations derived from known microphysics --- both have their place, but ultimately we aim to see why the former is a consequence of the latter.

The second is the empirical search for fundamental non-unitary (as occurs in dynamical collapse theories like the GRW theory, continuous state localization, or the Diosi-Penrose model; see, \egc, \cite{bassi-dorato-ulbricht} for an up-to-date review of extant collapse theories and their empirical status). Here it is essential (the first principle) to rule out the possibility that the non-unitarity is the consequence of interaction with an environment: only if this occurs do we have a genuine violation of quantum mechanics and not just emergent non-unitarity. But (the second principle) it is constitutive of `interaction with an environment' that the interaction can actually be modelled as the consequence of dynamical coupling with a second system. And (the third and fourth principles) evidence of fundamental non-unitarity does not require the experimenter to rule out \emph{every conceivable} dynamical interaction with another system (that would be in practice impossible) but only to rule out those dynamical interactions that our microphysical theories give us reason to expect to be present. 

(Of course, fundamental non-unitarity is new physics, and once we are considering new physics, other possibilities appear: perhaps the apparent non-unitarity is after all a consequence of entanglement with a hitherto-unsuspected new particle or field. No single experiment would suffice to falsify the whole quantum framework: the point is that one way or another such a discovery would point to new physics, but it would do so only because of the methodological commitment to attempt first to explain the observations in terms of known --- and unitary --- microphysical theories.)

Put aside whether the four principles correctly describe physics practice. Why think that they are correct?  The obvious answer is just that the strategy of accounting for the dynamics of quantum systems in this way has been spectacularly successful. We have extremely good evidence that, for instance:
\begin{itemize}
\item Where nuclear interactions and electromagnetic radiation can be neglected, systems are accurately described by nonrelativistic quantum mechanics.
\item Where nuclear interactions but not electromagnetic radiation can be neglected, systems are accurately described by quantum electrodynamics (and, in the nonrelativistic regime, by quantum optics).
\item Where effects at energy scales above 10 TeV can be neglected, systems are accurately described by the standard model.
\item Where Planck-scale physics can be neglected, self-gravitating systems (insofar as quantum effects are also present) are accurately described by low energy quantum gravity (cf \citeNP{wallaceleqg} and references therein).
\end{itemize}
And in each of these cases, the `can be neglected' clause is not empty tautology but can be quantitatively assessed (for instance, in terms of typical energy levels) and normally checked if necessary by evaluating leading-order corrections from a more fundamental description. In short, both the metaphysical and methodological versions of the four dogmas of quantum reductionism are justified by the evidence that we actually do know what the correct microphysical theories of the world are, outside some fairly exotic contexts, and that \emph{in fact} those theories are unitary quantum theories.

Furthermore, this `extremely good evidence' does not simply come from microphysical explorations: it does not, that is, just consist of evidence from particle accelerators and the like combined with an \emph{a priori} commitment to reductionism. It consists in large part of the thousands of successes of the overlapping fields of (equilibrium and non-equilibrium) quantum statistical mechanics, condensed matter physics, and effective field theory in deriving empirically-confirmed claims at the phenomenological level from quantum microphysics. We have not by any means explained \emph{every} quantitatively-known quantum dynamical system in this way, but we have explained a great many: to pick just a few examples, we have quantitatively-confirmed accounts of:
\begin{itemize}
\item the linear responses of many forms of matter to external perturbations (through the relation of linear-response-theory coefficients to two-time expectation values and the evaluation of the latter in perturbation theory; see \cite{zwanzig1965review} for a classic review);
\item the master equations governing atomic and nuclear decay of excited states;
\item the phenomena of superconductivity and superfluidity in a wide class of solids and fluids via the spontaneous breaking of (respectively) local and global phase symmetry;
\item the effective theory of pion scattering (via the identification of the pion with the Nambu-Goldstone boson associated to spontaneous breaking of chiral symmetry in quantum chromodynamics);
\item the suppression of quantum interference in mesoscopic systems through decoherence theory.
\end{itemize}

Reductionist claims of this modest kind are fairly uncontroversial in physics: Anderson, for instance, is often held up by philosophers as an arch-critic of reduction, but his famous~~\citeyear{andersonmoredifferent} discussion of the subject begins thus:
\begin{quote}
The reductionist hypothesis may still be a topic for controversy among philosophers, but among the great majority of active scientists I think it is accepted without question. The workings of our minds and bodies, and all the animate or inanimate matter of which we have any detailed knowledge, are assumed to be governed by the same set of fundamental laws, which except under certain extreme conditions we feel we know pretty well.
\end{quote}
But Anderson isn't wrong that they are controversial in philosophy of science. And notwithstanding the various quantitative successes of quantum statistical mechanics, both philosophers and physicists have claimed that various paradoxes of principle block its derivations: the measurement problem, the thermodynamic limit, the irreversibility problem of statistical mechanics. (I return to the latter in section~\ref{arrowoftime}.) Defending the thesis systematically is far more than I can do here; I will settle for the weaker claims that (a) a methodological commitment to reductionism in this sense has been intimately entangled with quantum theory itself since its inception (and indeed before), and (b) that this methodology has been extremely successful scientifically. 

Although my focus here is on quantum theory, modern physics' use of \emph{classical} physics adheres to a modified version of the same four dogmas. Here too there is a plurality of dynamics, with many classical systems obeying stochastic dynamics (like the Langevin equation) and/or dissipative dynamics (like the Navier-Stokes equation) instead of Hamiltonian dynamics (the classical analog of unitary dynamics). But physics is methodologically committed (1) to explaining non-Hamiltonian dynamics by dilation to larger systems, insofar as it can be explained classically at all, and (ii) to model those larger systems via Hamiltonian dynamics. The difference is that in classical physics there is always the possibility of a \emph{quantum} explanation of the form of a classical equation, and there is no simple, short list of classical microphysical theories that suffice to explain all of classical dynamics: unification to that degree requires quantum mechanics. For more on this point, see \cite[pp.196-7]{wilsonwandering} and \cite[section 4]{wallacecategory}.

\section{The Open Systems View: a comparison}\label{opensystems}

Cuffaro and Hartmann~(\citeyearNP{cuffarohartmann}; henceforth CH) have recently expounded and defended what they call the `open systems view' of quantum mechanics, according to which we should take open, rather than closed, quantum systems as fundamental and treat closed systems as a special case of open systems. The Open Systems View clearly has a lot in common with the view I have been developing and advocating; in this section I want to consider how the two are related.

Firstly: what exactly is the Open Systems View? CH give two characterizations, which are somewhat subtly related. The first is a \emph{physical characterization}: an open system is one which is dynamically coupled to its environment, while a closed system is isolated from it. (Open systems are ``systems evolving under the influence of their environment''.) The second is a \emph{formal characterization}: an open quantum system is a quantum system with dynamics that is linear and preserves the consistency of the Born probability rule but need not be unitary; a closed quantum system has unitary dynamics. (CH are mostly interested in Markovian open systems, but I don't think the Markovian restriction is essential to them.)

Thus stated, the Open Systems View seems compatible with the present view, but it contains an important tacit assumption worth making explicit: the `open systems' CH consider are required to be \emph{autonomous} systems. In general, as we have seen, `systems evolving under the influence of their environment' might have no self-contained dynamics at all, but CH are specifically concerned with systems that do have well-defined dynamics; they simply want to drop the assumption of unitarity.

Once this is recognized, one of CH's original arguments for the Open Systems View loses its force: they argue that `there are reasons stemming from physics to believe that the cosmos is the \emph{only} closed system that truly exist', because no system other than (perhaps) the cosmos can be entirely isolated from its environment. But strictly speaking, no system has exactly autonomous dynamics either: autonomy, like isolation, is always an idealization, and manifestly there are many systems studied in quantum theory which are satisfactorily idealized as isolated (and unitary).

This is not, however, a serious problem for CH. What matters in their argument, and what does the bulk of the work in their paper, is the physical argument that there are plentiful examples of autonomous quantum systems that are not isolated, and the formal argument that subsystems of unitary quantum systems can display autonomous but non-unitary dynamics. I endorse both claims in the current paper (indeed, CH's formal argument is closely related to my argument in section~\ref{open-system-derivation}, though CH are less concerned with the mechanisms that ensure autonomy); thus far, the Open Systems View seems fully compatible with the present view. (And CH  provide an invaluable explication and defense of some of the metaphysics of open systems, and its interaction with issues like the quantum measurement problem, most which is applicable here also.)

Looking at the details of CH's view, however, reveals very substantial points of disagreement, which can be summarized as saying that CH appear to reject most or all of the four principles of quantum reductionism. 

The first principle is that non-unitarity implies interaction. CH are equivocal about this principle: they note (section 5.3) that non-unitary terms in a system's dynamics `are most naturally interpreted as due to the system's interacton with an external system', but are willing to countenance non-unitary dynamics for the cosmos as a whole and in the context of dynamical-collapse theories. (As an aside: as we have seen, renormalization means that we should not think of interactions as simply introducing non-unitarity: some part of a system's unitary dynamics might also be explained by its environment.)

CH \emph{seem} to deny the second principle, that interaction is derived from dilation. (And insofar as they do, they effectively deny the third and fourth principles, which presume the second.) They say explicitly (section 2.2) that 
\begin{quote}[u]nlike [unitary quantum mechanics]. for which the dynamics of an open system of interest, \mc{S}, is obtained via a contraction \ldots of the total dynamics of \mc{S}+\mc{E} to the state space of \mc{S}; in [the Open Systems View], the dynamical equations that govern the evolution of \mc{S} pertain directly to \mc{S} itself. Systems are modelled as genuinely open\ldots we do not describe the influence of the environment on \mc{S} in terms of an interaction between two systems, but instead represent the environment's influence on the dynamical equations that we take to govern the evolution of \mc{S}.\end{quote}

I quote at length because this seems to encapsulate much of the essence of CH's view, and also because I find it somewhat difficult to interpret. CH seem to be saying that a dynamical equation describing the evolution of \mc{S} in the presence of its environment `pertains directly to \mc{S} itself' if it is posited directly, but not if it is derived from contraction of the dynamics of \mc{S}+\mc{E}. But at the formal level, the dynamics of \mc{S} pertains to \mc{S} simply by virtue of \emph{being} an autonomous dynamics for \mc{S}: that is, being a rule that determines the future state of \mc{S} from its present and past states, perhaps given some broad-brush information (e.g. temperature) about \mc{E} but without regard to \mc{E}'s fine-grained state. And at the physical level, surely what it means for a term in the equation to pertain to \mc{S}+\mc{E} is that it represents the effect of \mc{E} on \mc{S}: if, physically, some interaction pertains only to \mc{S}, it's difficult to see how it can represent the effects of something other than \mc{S}.

More seriously: if we drop the idea that the effects of an environment on a system can be analyzed in terms of dynamical couplings between the two, it becomes difficult --- practically but also conceptually --- to distinguish intrinsic non-unitarity from the effects of the environment, and to understand just how it is that the environment influences the system. To say that a system is influenced by the environment is normally to say something, even if schematically, about how that influence occurs and how it might be manipulated, and saying that in quantum theory seems to require us to model the dynamical interaction between the two --- which is what CH seem to say we should not do. 

In quantum computation, for instance, it is of great importance to minimize the effects of environment on system. Doing so requires us to understand the physics of those effects, so as to manipulate them: we want to know, for instance, whether and how environment-induced non-unitarity depends on environmental temperature, and whether various forms of screening might reduce the environment's effects. Doing so --- at least in the standard practice of physics, but arguably also in principle --- requires us to model system plus environment together, so that the induced dynamics of the system can be related to the environment's macrostate and the form of the interaction between the two. (It matters, of course, that the practice of modelling system-environment interactions in this way is highly successful: physics goes far beyond the phenomenological here.)

Similarly, in testing dynamical-collapse theories, we have seen already that the name of the game is to identify and eliminate environment-induced decoherence, so that any residual non-unitarity must be regarded as intrinsic. If the effects of the environment are not to be analyzed in the usual way via dynamical interactions, it is opaque --- to me at any rate --- how even in principle we could obtain evidence for intrinsic non-unitarity.

Finally, there is a question of self-consistency for the Open Systems View thus understood. Suppose $\mc{S}$ and $\mc{T}$ are both open systems. There is a system $\mc{S}+\mc{T}$ that comprises both $\mc{S}$ and $\mc{T}$; it, too, presumably is an open system. The dynamics for $\mc{S}$ alone, and $\mc{T}$ alone, will be derivable from the joint dynamics of $\mc{S}$ and $\mc{T}$, and that derivation will reveal the dynamics of $\mc{S}$ and $\mc{T}$ separately to be, after all, derived from contraction of the joint dynamics of $\mc{S}+\mc{T}$.

For these reasons, I doubt that it is viable for the Open Systems View to reject the second principle --- and, partly because of this, I am unclear as to whether CH intend to do so. If after all we interpret them as \emph{accepting} the second principle, but continue to take them seriously about the inessential role played by unitary dynamics in particular, then we would obtain a coherent version of the Open Systems View which rejects the third principle (that dilations are always to larger, closed systems).
In a view like this, we can always dilate a system in order to study its interactions with its environment, but the dilated system in general will itself be an open system --- there is no expectation that eventually this process terminates in a closed (i.e., unitarily-evolving) system.

A view of this kind could be combined either with the denial or the acceptance of the fourth principle (that the systems to which we dilate are, outside exotic situations, governed by known dynamical laws. If the fourth principle is denied, we end up with an anti-reductionist position that nonetheless has a consistent methodology for scientifically investigating the dynamical causes of non-unitarity in any given system. If it is accepted, we get a view where the fully dynamics of any given system is a combination of known and unitary microdynamics plus a non-unitary interaction term; the latter term can be analyzed by dilation, but only to a system which in turn has non-unitary interactions, with the whole process continuing without limit.

Both of these views seem coherent. The problem with either is inductive: \emph{in fact} physics, most notably in the (overlapping) realms of statistical mechanics, condensed matter, and effective field theory, has had remarkable success explaining the quantitative dynamics of open quantum systems by dilation to larger \emph{closed} systems, and by assuming that at sufficiently short lengthscales systems are governed by unitary, and largely known, dynamics. This both provides support for the methodological validity of all four principles, and would be difficult to explain without their metaphysical validity.

(At this point I should consider CH's main physics example of how assuming the primacy of unitary dynamics supposedly leads to trouble: the evaporation of black holes. \citeN{hawking1975} proposed that the evaporation process was non-unitary; CH observe that this is an entirely legitimate possibility on the Open Systems View and that (section 1) ``at least part of the motivation for wanting to reject it seems to amount to nothing more than that it runs counter to quantum theory and the fundamental dynamics described by it [on a Closed Systems view]''. 

At some level this is a semantic dispute: we could take intrinsic (\iec, not environment-induced) non-unitarity to be included in quantum mechanics, or we could take it as a modification of quantum mechanics. (On either approach, \emph{experimental} evidence for intrinsic non-unitarity would require prior elimination of environmental factors, but that does not apply in this theoretical context.) What matters is whether physicists were disinclined to consider Hawking's proposal due in part to an in-principle rejection of non-unitary dynamics, and it is not at all clear to me that this is the case --- of course many, many people have commented on the information-loss paradox, and some of those comments (especially in popular and semi-popular articles and books) have claimed a contradiction with quantum mechanics, but more detailed and technical arguments seem to have carried more weight --- notably Banks, Peskin and Susskind's \citeyear{bankspeskinsusskind} argument that non-unitary effects would be magnified to empirically-ruled-out scales by loop corrections in quantum field theory, and more decisively, the evidence for AdS/CFT duality (\citeNP{maldacenaconjecture,wittenadscft}; cf \citeN{wallaceinformationloss} and references therein for a conceptual discussion), which if correct shows that black hole evaporation is dual to a manifestly unitary process.)

The choice between quantum reductionism and the Open Systems View is largely empirical, and rests on whether to accept my claim that reductionism (in the weakish sense I intend) is indeed underwritten by the successes of physics. Even if this claim is granted, though, it has one interesting exception, one place where even a reductionist analysis of open quantum systems potentially needs resources not available in unitary quantum mechanics: the problem of the arrow of time in non-equilibrium statistical mechanics.

\section{The arrow of time}\label{arrowoftime}

Unitary dynamics is reversible: the Schr\"{o}dinger equation can be run backwards in time as easily as forwards, and there is no intrinsic difference between the two. This is not true for non-unitary dynamics: an autonomous but non-unitary dynamics is irreversible, in the sense that two distinct density operators evolving under that dynamics approach one another in trace norm. Physically (and assuming a dilation-based description of the system-environment interaction) this corresponds to information loss: the system gets entangled with the environment; taking the partial trace discards this information. In any case, the future direction in time is picked out as the direction of information loss and of convergence in trace norm.\footnote{Note that this distinction is independent of whether the unitary dynamics being considered is time-reversal-invariant in the usual sense: even time-reversal-noninvariant unitary processes are reversible. See \cite{wallacereversalreversibility} for more on this point.}

It is then in principle impossible to find an \emph{exceptionless} derivation of autonomous dynamics for \mc{S} from unitary dynamics for $\mc{S}+\mc{E}$, except in the special case where $\mc{S}$ is not just autonomous but itself unitary. And so some initial assumption must have been made in the ``derivation'' sketched in section~\ref{open-system-derivation}. It is easy to identify that assumption: I assumed that the third term in the Nakajima-Zwanzig equation (\ref{exactexpression}) vanished, something I argued for on the grounds that the microscopic degrees of freedom of the environment ``effectively equilibrated''. Of course,  equilibration is itself a time-directed process, and the assumption that a given state equilibrates amounts to the assumption that the state lacks the delicate correlations that would prevent equilibration (see \citeN{wallacelogic} for more on this point). 

The conclusion of all this is that deriving autonomous dynamics from unitarity requires an assumption about the initial state of system-plus-environment --- a harmless-seeming, intuitively-reasonable assumption, perhaps, but an assumption which holds for \emph{initial} states but on pain of contradiction cannot hold in time-reversed form for \emph{final} states. In most approaches to the foundations of statistical mechanics, it is this initial-state assumption that underpins irreversibility and which must be explained to explain irreversibility: proposed explanations have included an operationalist (and itself inherently time-directed) requirement that systems must always be thought of as prepared by an agent who is unable to prepare them in sufficiently complicated states to prevent effective equilibration~\cite{myrvoldbook}, a (time-directed) principle of causal reasoning that legitimates certain no-correlation assumptions for initial but not final states~\cite{maudlinobjective}, and a cosmological boundary condition of one kind or another~\cite{Penrose1994,alberttimechance,wallacelogic}.

There is, however, another tradition in foundations of statistical mechanics: the \emph{interventionist} tradition which seeks to explain irreversibility by the effects of an external environment. Applied to our combined system $\mc{S}+\mc{E}$, an advocate of this tradition would point out that (i) realistically $\mc{S}+\mc{E}$ is itself not completely isolated, and (ii) adding even a small amount of non-unitarity to the combined dynamics of $\mc{S}+\mc{E}$ probably suffices to destroy the delicate correlations that would allow any given initial state \emph{not} to proceed effectively to equilibrium with respect to the microscopic degrees of freedom of $\mc{E}$.

The majority view in foundations of statistical mechanics (defended explicitly by, \egc, \citeN[pp.250-254]{sklarstatmech}) is that interventionism is not the right way to explain irreversibility, in part not because the argument is incorrect in itself but because it just pushes the problem of explaining irreversibility back one step, leaving unanswered the \emph{general} question of why systems are so aligned as to produce a consistent direction of irreversibility. I will not attempt to resolve this controversy here. But it is worth observing that environment-based accounts of irreversibility fit very naturally into the Open Systems View --- especially into the rather modest version of that view I discussed at the end of section~\ref{opensystems}, which endorsed the first, second and fourth principles of quantum reductionism while dropping the requirement that autonomous dynamics are always explainable via dilation to a \emph{closed} system. Here at least, the advocate of an Open Systems View has a potential rejoinder to the inductive argument based on the success of closed-system, reductive analyses of autonomous dynamics.

\section{Conclusions}

If quantum mechanics were to be understood first and foremost as an exactly-true theory of the universe, with its applications to subsystems and to restricted scales understood as derivative or instrumental, then it would be natural to assume that unitarity, and the use of pure rather than mixed states, were fundamental. But there is no reason to do this: the Fundamentality Assumption (that physical theories should be understood as if they were exactly true) and the Cosmological Assumption (that physical theories should be seen as modeling the whole Universe) have no justification in scientific practice or naturalized epistemology. Outside a few explicitly cosmological contexts, our goal in physics (and specifically in quantum physics) is in general to model the dynamics of autonomous, though not necessarily isolated, subsystems of the larger Universe.

Thus understood, quantum mechanics inevitably leads to the conclusion that quantum states need not be pure and that quantum dynamics need not be unitary. Quantum statistical mechanics provides many examples in which a subsystem of a larger unitary system displays autonomous dynamics not because its degrees of freedom are simply uncoupled to those of the rest of the system but because scale separation and effective equilibration of the system's environment leads to autonomous, but non-unitary, dynamics for the subsystem.

This does not, however, imply a simple pluralism of dynamics in which unitary evolution has no special place. Physics is committed, and justifiably so, to four broadly-reductionist principles that can be seen as a naturalized replacement for the unjustified Fundamentality and Cosmological Assumptions, namely (i) that non-unitary dynamics should be explained (except as a last resort) by interaction with an environment; (ii) that `interaction with an environment' should be mathematically modelled through dilation to a larger quantum system; (iii) that for the purposes of modelling an autonomous subsystem it always suffices to dilate to a \emph{unitary} system; (iv) that we should seek to analyze any such system, ultimately, in terms of the known, and unitary, dynamics of microscopic physics: nonrelativistic particle mechanics, QED, the Standard Model, and so forth. (i) is largely justified methodologically (non-unitarity is evidence for \emph{intrinsic} non-unitarity only once interaction with an environment is ruled out). (ii) is constitutive of what it means for a term in the dynamics of a system to represent interaction with other systems. (iii) and (iv) are not true \emph{a priori}, but receive very strong evidential support from the manifold successes of quantum mechanics, especially in statistical mechanics, condensed-matter physics, and effective field theory.

 Cuffaro and Hartmann, in developing their Open Systems View, also reject the Fundamentality and Cosmological assumptions, and also center the principle that autonomous quantum systems often evolve non-unitarily. Their view differs from the view here by its rejection of at least some of these four principles of quantum reductionism, though it is to some degree a matter of interpretation just which principles they wish to reject. At least on my reconstruction of their views, it is a semantic matter whether or not they accept the first principle, and a methodological necessity for them to accept the second principle. The status of the fourth principle depends on one's assessment of the success of reductionist methods (in the moderate sense I intend for `reductionist') in modern quantum mechanics. The third principle is most interesting: its acceptance or rejection is tied to how the problem of irreversibility in statistical mechanics is to be solved.

\section*{Acknowledgements}

Thanks to Michael Cuffaro and Stephan Hartmann for the stimulating work that inspired this paper, and for helpful correspondence.


\end{document}